\documentclass[lettersize,journal]{IEEEtran}
\usepackage{amsmath,amsfonts}
\usepackage{algorithmic}
\usepackage{algorithm}
\usepackage{multirow}
\usepackage{makecell}
\usepackage{array}
\usepackage[caption=false,font=normalsize,labelfont=sf,textfont=sf]{subfig}
\usepackage{textcomp}
\usepackage{stfloats}
\usepackage{url}
\usepackage{verbatim}
\usepackage{enumerate}
\usepackage{graphicx}
\usepackage{threeparttable}
\usepackage{cite}
\DeclareUnicodeCharacter{2212}{-}
\begin{document}

\title{An Ontology-based Method to Identify Triggering Conditions for Perception Insufficiency of Autonomous Vehicles}

\author{Xingyu Xing, Tong Jia, Junyi Chen, Lu Xiong, Zhuoping Yu
\thanks{This work was supported in part by the National Key Research and Development Program of China under Grant 2021YFB2501205. \textit{(Corresponding author: Junyi Chen)}}       
\thanks{All authors are with the School of Automotive Studies, and also with the Institute of Intelligent Vehicles, Tongji University, Shanghai 201804, China. (e-mail:xingxingyu,jiatong,chenjunyi,xiong\_lu,yuzhuoping@tongji.edu.cn).}
}

\markboth{Journal of \LaTeX\ Class Files,~Vol.~14, No.~8, August~2021}%
{Shell \MakeLowercase{\textit{et al.}}: A Sample Article Using IEEEtran.cls for IEEE Journals}


\maketitle

\begin{abstract}
The autonomous vehicle (AV) is a safety-critical system relying on complex sensors and algorithms. The AV may confront risk conditions if these sensors and algorithms misunderstand the environment and situation, even though all components are fault-free. The ISO 21448 defined the safety of the intended functionality (SOTIF), aiming to enhance the AV's safety by specifying AV's development and validation process. As required in the ISO 21448, the triggering conditions, which may lead to the vehicle's functional insufficiencies, should be analyzed and verified. However, there is not yet a method to realize a comprehensive and systematic identification of triggering conditions so far. This paper proposed an analysis framework of triggering conditions for the perception system based on the propagation chain of events model, which consists of triggering source, influenced perception stage, and triggering effect. According to the analysis framework, ontologies of triggering source and perception stage were constructed, and the relationships between concepts in ontologies are defined. According to these ontologies, triggering conditions can be generated comprehensively and systematically. The proposed method was applied on an L3 autonomous vehicle, and 20 from 87 triggering conditions identified were tested in the field, among which eight triggering conditions triggered risky behaviors of the vehicle.
\end{abstract}


\section{Introduction}
\IEEEPARstart{T}{he} autonomous vehicle (AV) is a typical complex system that includes the environment perception module, decision-making module, and motion control module. The AV's safety depends on each module's good performance and fail-safety strategy when malfunctions of electronic and electrical components occur, which involves functional safety\cite{ref1}. Meanwhile, since self-driving implementation relies on the accurate perception and correct understanding of the environment through sensors and algorithms, AVs confront the challenge from the safety of the intended functionality (SOTIF). That means AVs may still enter a hazardous situation due to unintended behaviors when parts and modules function well. Aiming at this issue, international organizations in the industry are committed to developing relevant standards and methods. The first version standard for SOTIF, ISO/PAS 21448, was released in 2019\cite{ref2}. The standard proposes a development process to ensure SOTIF and provides a testing and validation framework. Its content includes safety analysis for AVs, modifying and optimizing the intended functions, defining the validation strategies, evaluating known hazardous scenarios, and exploring and testing unknown hazardous scenarios. Its goal is to standardize the process of product development and validation in order to provide reliable evidence for the system's safety argumentation. As for the safety analysis, system hazards need to be identified and assessed, which is primarily similar to ISO 26262. After finishing the safety analysis, potential functional insufficiencies and triggering conditions should be identified.

Triggering conditions are specific conditions of a scenario\cite{ref2} that serve as an initiator for a subsequent system reaction leading to hazardous behavior, such as harsh environmental conditions or rare targets. When the AV has certain functional insufficiencies and performance limitations, under the influence of triggering conditions, the vehicle will potentially enter a dangerous situation due to the system's unexpected behavior deviating from the intended function. Therefore, it is necessary to identify and analyze the triggering conditions and verify the system's behavior under the triggering conditions.

The performance limitations of the perception system and related influencing factors will make it produce unsatisfying perception results, which severely impact the AV's safety. Typical environmental perception sensors used in AVs can be divided into active and passive sensors. These sensors rely on physical principles like light and electromagnetic wave propagation. Scholars have carried out some research on this issue about different kinds of sensors. Ian Colwell et al.\cite{ref3} introduced the perception triangle and defined the perception uncertainty as sensor properties, labeling uncertainty, model uncertainty, and so on. Lei Ren et al.\cite{ref4} researched the influence of different environmental conditions such as weak lightness, rain and snow, lens blur, and image rotation on camera perception and recognition algorithms. OpenCV was used to process images from the KITTI dataset to generate images of different environmental conditions. The average precision (AP) of vehicle detection was used as the evaluation index. The results showed that the extreme environment has a significant influence on the recognition algorithms. Sinan Hasirlioglu et al.\cite{ref5},\cite{ref6} used smoke simulators and rain simulators to quantitatively study the effect of fog and rain on the Camera. The results show the decrease of image contrast with the increase of density of fog and rain. Liang Peng\cite{ref7} studied the performance of image recognition algorithms under uncertainty coming from extreme weather and adverse lighting and proposed a Monte Carlo dropout method to analyze the uncertainty. Philipp Rosenberger et al.\cite{ref8} proposed a LiDAR model containing three categories of influencing factors: medium transmittance, reflection characteristics of the object surface, and illumination conditions; A. Filgueira et al.\cite{ref9} experimented with the effects of rainfall on the amount and intensity of LiDAR's point cloud and ranging accuracy. David McKnight et al. \cite{ref10} further studied the impact of precipitation on LiDAR by comparing experimental results with several attenuation models and verified that fog and rain would dramatically decrease the maximum range of measurement. All these related works proved the influence of several common external factors on the perception system through experimental or theoretical approaches and gave some valuable suggestions to improve perception performances. However, for the purpose of supporting triggering condition analysis in the SOTIF process, it needs a relatively complete and systematic analysis method to identify all kinds of factors for safety verification. The ISO 21448 provides some requirements and recommendations for triggering condition analysis but lacks specific approaches. Aiming to solve this issue, an edge case generation method based on expert knowledge and theoretical analysis was proposed in our previous studies\cite{ref11}. A list of factors as complete as possible based on theoretical analysis of the physical and algorithm principle of perception systems was formed, then generated edge cases at the semantic level. In another research, a framework to support safety analysis and verification of AVs based on triggering conditions\cite{ref12} was introduced.

In this paper, an analysis method of perception insufficiencies and triggering conditions of the AV is proposed. Based on the working principle of the perception system, an analysis framework of triggering conditions based on the propagation chain of events model for perception errors, consisting of triggering source, perception process, and triggering effect, is introduced. Ontologies of triggering source and perception stage are constructed, and the relationships between concepts in ontologies are defined. According to these ontologies, triggering conditions can be generated comprehensively and systematically. The main contributions of this paper are as follows:

\begin{itemize}
\item{The propagation chain of events model is applied to describe the perception error resulting from triggering conditions. An analysis framework was proposed based on the propagation chain of events model. The framework consists of the triggering source where the influencing factor comes from, the perception process from sensing to recognition, and the triggering effect which the triggering source leads to. The framework can support identifying triggering conditions related to the perception system.}
\item{An ontology-based method to generate triggering conditions was introduced. The categories of triggering conditions cover the physical properties of precepted targets, different environmental conditions, relationships of environmental elements. Semantic test cases combining triggering conditions and traffic scenarios can be generated to support the verification process that complies with ISO 21448.}
\item{An L3 autonomous vehicle was analyzed and tested using the introduced method, proving the method to be valid. Compared with methods that depend on expertise and experiences, this method provides a more formal and structured way to describe triggering sources and identify triggering conditions more wholly and systematically.}
\end{itemize}

This paper proceeds as follows. Section II introduces the propagation chain of events model of perception error led by triggering conditions and the analysis framework. Section III introduces the ontology structure and relationships of triggering sources and the perception stages. Based on the ontology, the method to generate specific triggering conditions is proposed. In section IV, an application of the method in the verification process of a self-driving sweeping vehicle will be given. Section V is the discussion, and section VI is the conclusion and some future works.

\section{Analysis framework of triggering conditions}
\begin{figure}[!htpb]
\centering
\includegraphics[width=3.2in]{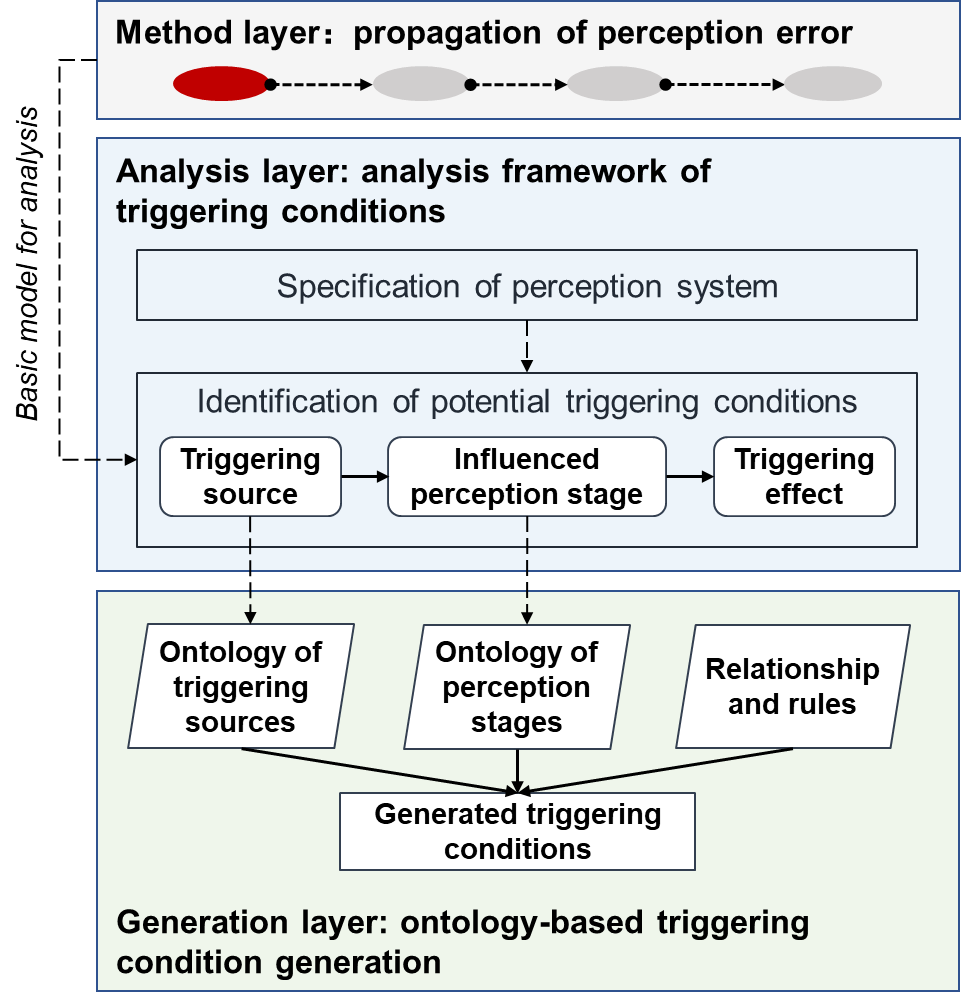}
\caption{Framework for analysis and generation of triggering conditions}
\label{Figure 1}
\end{figure}

According to ISO/PAS 21448, triggering conditions should be analyzed and evaluated. Then, the vehicle's behavior under triggering conditions needs to be verified with hazardous scenarios. In this section, the propagation of perception error which provides the basic model for triggering condition analysis is introduced. Based on the propagation chain of events, the analysis framework consisting of the triggering source, influenced perception stage, and the triggering effect is proposed. Furthermore, to support an effective analysis and verification of triggering conditions, an ontology-based method to generate triggering conditions is proposed in section III. The relationship between the analysis framework and the ontology-based generation method is shown in Fig.\ref{Figure 1}. 

\subsection{Propagation of perception errors resulting from triggering conditions}
\begin{figure*}[!htpb]
\centering
\includegraphics[width=7.16in]{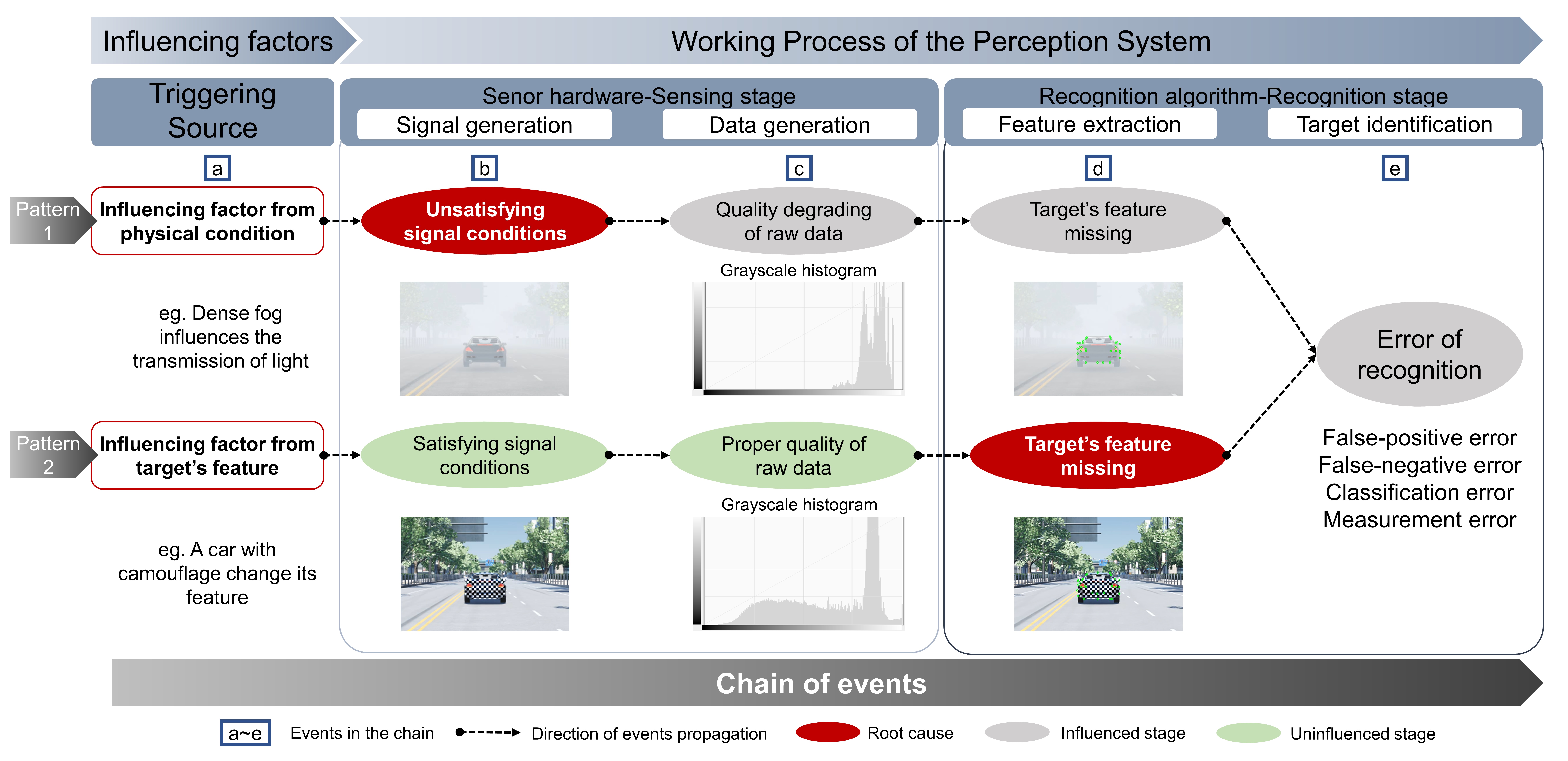}
\caption{Propagation chain of events for the triggering condition causing perception error.In pattern 1, the root cause is the worst-cases of physical conditions. It leads to the target's features missing and error of recognition. In pattern 2, the root cause is the worst-cases of the target's features. }
\label{Figure 2}
\end{figure*}

Although plenty of research and experiments have demonstrated that the environmental influencing factors substantially impact perception performance, only influencing factors that lead to perception error and system hazard are regarded as triggering conditions. To analyze the potential triggering conditions systematically, we use the propagation chain of events model to describe the formation of perception errors resulting from triggering conditions. In the chain of events model, each event propagates in a direct causal manner to the next event in the linear time-ordered sequence\cite{ref13}. Because inside the perception system, the information processing procedure is sequential and has no interaction with outside elements, the conventional chain of events model is suitable to portray this process. In this paper, the propagation chain of events is constructed based on the working process of the perception system, as Fig.\ref{Figure 2} shows. We divided the perception process into the sensing and recognition stages. In the sensing stage, raw data of the environment, for example, the image and point cloud, is acquired by sensor hardware. In the recognition stage, information about targets is identified by the recognition algorithm from the raw data. We furtherly decomposed each stage into sub-stages. The sensing stage consists of signal generation relying on physical principles and data generation relying on chips and algorithms. The recognition stage consists of feature extraction and target identification.

According to the propagation model, any unsatisfying result in sub-stages mentioned above can become a root cause and conduct to the subsequent sub-stage, leading to a perception error. For SOTIF involved perception error, these unsatisfying results come from environmental influencing factors, which are regarded as triggering conditions. In order to construct environmental influencing factors into the propagation, we define the concept of triggering source. Triggering sources are the environmental elements that contain the influencing factors and have the potential to form triggering conditions. Moreover, different triggering sources will produce different triggering effects on the related perception stage. To describe the triggering condition completely, we need to specify the triggering source, the influenced perception stage, and the corresponding triggering effect. For example, according to the accident report of a Tesla car in 2016\cite{ref14}, the camera system did not recognize the tractor-trailer. In this case, the triggering source is the white tractor-trailer against the bright lit sky; the influenced perception stage is feature extraction in the recognition stage; the triggering effect is the failure of feature extraction and the false-negative error. 

There are two patterns of perception errors depending on the source of root causes and their affected perception stages: physical-condition-based and target-feature-based. The propagation of a perception error resulting from the physical-condition-based triggering conditions is as follows.

\begin{enumerate}[a)]
\item{\textbf{\textit{Influencing factors from physical conditions:}} These influencing factors come from the triggering source's properties related to the sensor's working principle, For Instance, the triggering source is dense fog—the refraction property of fog influences the transmission of light.}
\item{\textbf{\textit{Unsatisfying signal conditions:} }They are the root causes of physical-condition-based perception errors. Affected by influencing factors from the triggering source, the sensor generates unsatisfying signals. When the dissatisfaction exceeds the sensor's ability to deal with, environment information will be missed by the sensor partially.}
\item{\textbf{\textit{Quality degrading of raw data:}} Due to the unsatisfying signals, the quality of raw data drops seriously. In the example, it is the improper brightness and contrast of the perceived image. For LiDAR, it may be the points cloud missing or the ranging accuracy reducing.}
\item{\textbf{\textit{Target's feature missing: }}Due to the decreased quality of raw data, some necessary features to recognize the target are lost or distorted, making it difficult for the algorithm to extract.}
\item{\textbf{\textit{Recognition error:}} The algorithm cannot recognize the target because of the loss of features, leading to false-positive, false-negative recognition results or classification and measurement errors.}
\end{enumerate}

For the target-feature-based perception error, the sensing stage is hardly influenced by physical conditions, and the signal conditions and perceived raw data are satisfying for the recognition algorithm, while the root causes lay in the target's features. The disappearance of features may be due to the similarity of targets and background, or the occlusion of targets. Even though features are clearly extracted, the algorithm still may identify the target as the wrong category.

\subsection{Analysis framework based on the propagation of perception errors}
\begin{figure}[!bp]
\centering
\includegraphics[width=3.2in]{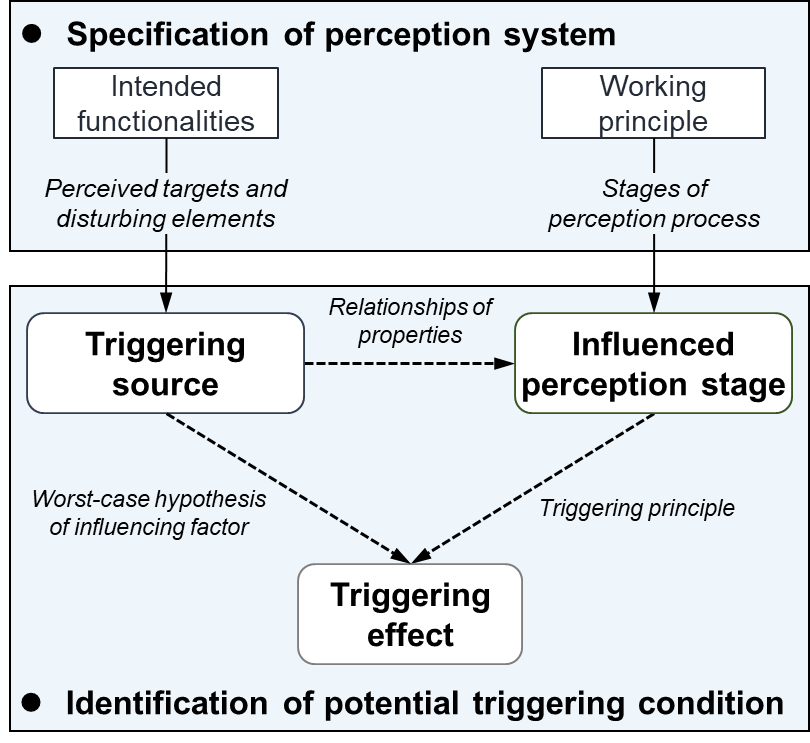}
\caption{Identification process of triggering conditions according to the propagation chain of events. The triggering principle is how triggering sources affect the perception system and lead to triggering effects. The principle is formed by analyzing and summarizing the working principle of the perception system. Triggering effects are the adverse effects of triggering sources on the perception system by the triggering principle.}
\label{Figure 3}
\end{figure}

According to the propagation of perception errors abovementioned, we proposed the analysis framework to identify potential triggering conditions, as presented in Fig.\ref{Figure 3}. 

\subsubsection{Specification of the perception system}
In the beginning, the perception system under analysis should be specified, including its intended functionalities and working principle. Different perception systems will be influenced by triggering sources' different properties. Triggering sources can be any element in the environment, including the static and dynamic targets, atmospheric elements like weather, illumination, or particles. According to the intended functionalities and verification requirement, the specific triggering source under analysis is determined.

Meanwhile, based on the working principle of the perception system, the working process is carried out, and the perception stage affected by triggering sources should be identified. Taking the LiDAR as an example of the active sensor, it is a kind of radar system which can scan and obtain information by scattering laser and detecting the reflected signal. According to the working principle of LiDAR, the perception stage that can be affected include the laser propagation process, the laser reflection process of objects, and the process of the sensor receiving the reflected laser. First of all, the sensing stage or recognition stage that can be affected should be determined according to the root causes of two patterns of perception error. Then, a more detailed substage should be identified in the following analysis.

\subsubsection{Identification of potential triggering condition}
\begin{figure*}[!htpb]
\centering
\includegraphics[width=6.5in]{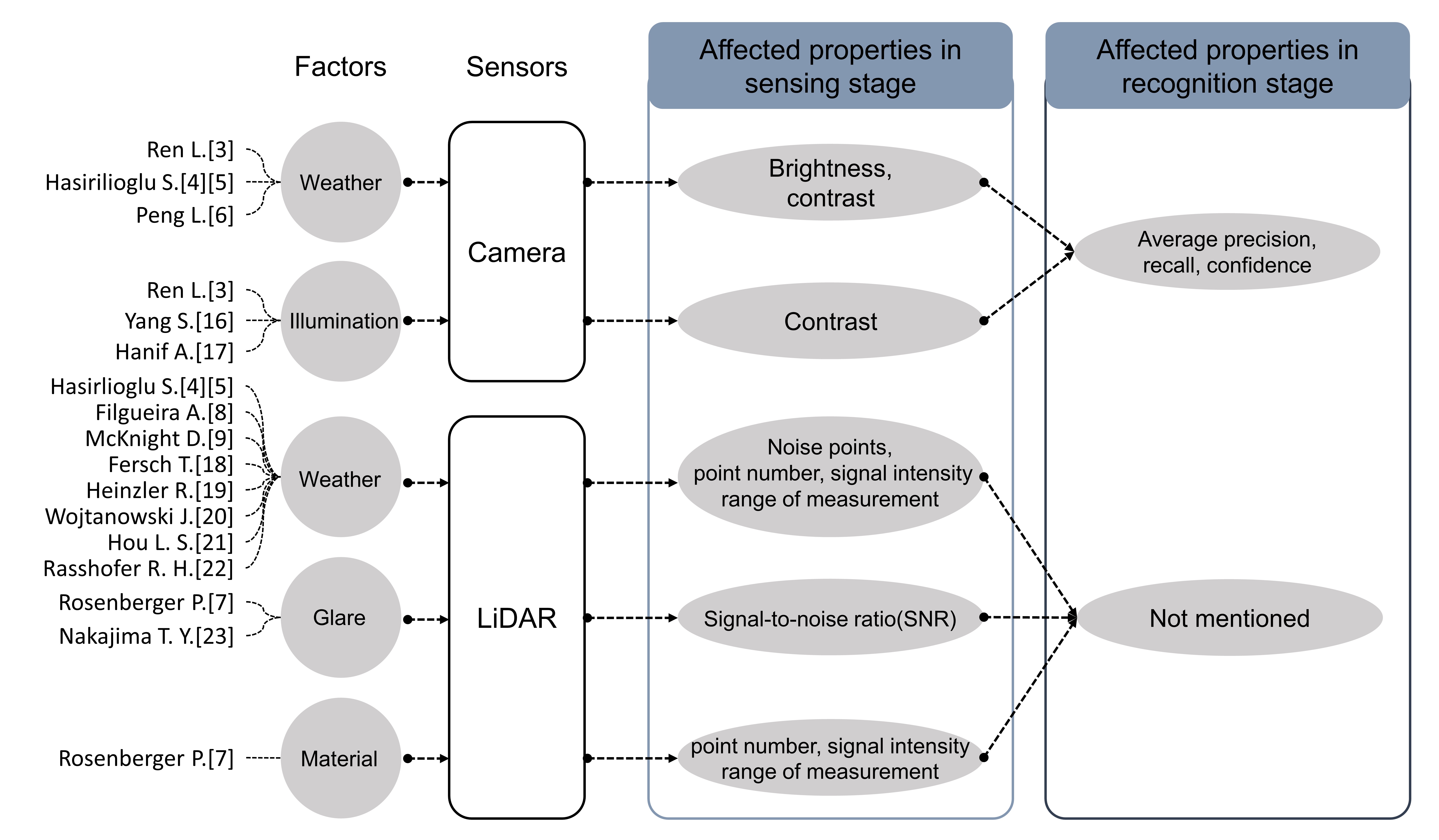}
\caption{Known influence factors and their affected properties of perception systems. Influencing factors studied in open literature include illumination, weather, glare, and material. For the camera, the mainly affected properties are brightness and contrast, and the average precision, recall, and confidence are used to study the influence on recognition results. For LiDAR, the mainly affected properties are point number, signal intensity, noise points, SNR, and range of measurement in the sensing stage.}

\label{Figure 4}
\end{figure*}

In order to identify potential triggering conditions, we should identify all the elements that constitute the triggering condition following the propagation chain of events model mentioned above. The triggering source and the affected perception stage are connected by the relationships of properties, which can be decided according to open studies or theoretical analysis based on the working principle of the perception system. For example, various particles in the atmosphere will interfere with the laser in the laser propagation stage. Large particles also reflect laser to the sensor, creating noisy laser points, such as noise from rain or snow. In the laser reflection stage of objects, the target surface's reflection characteristic, which is related to the material's transparency, color, and roughness, affects the reflected signal and leads to systematic deviation\cite{ref15}. Besides, the LiDAR's performance is limited by its angular resolution and installation position, which leads to the inability to perceive small obstacles and potholes in the distance\cite{ref16}. Fig.\ref{Figure 4} represents some known influencing factors and the affected perception stages.

It is impractical to take a theoretical analysis of the recognition algorithm. However, regardless of the unexplainable problem of algorithms, the core part of recognition is the extraction of features\cite{ref25}. Typical image features include texture, color, shape, and spatial relationship. Similarly, point-cloud-based target recognition requires clustering of points and geometric-model-based features. Therefore, the influencing factors of the recognition process mainly come from the variation of features of the perceived target. When the object's features are missing for various reasons, the algorithm will not be able to identify the target correctly. 

For the analysis of triggering effects, we mainly considered the triggering principle and the degree of influencing factors. The triggering principle describes the production of triggering effects relying on properties of triggering sources and the perception working process. For example, the triggering effect missing of the target's point cloud may occur due to a triggering principle of low reflectivity or a small reflection area. Another aspect that needs to be concerned is the degree of influencing factor. We introduced the worst-case hypothesis of influencing factors when analyzing possible triggering effects. Every perception sensor has a range of conditions to work effectively. Besides, considering the robustness and the fault-tolerant ability of the autonomous driving system, the subpar performance is not enough to cause serious adverse consequences. The worst-case hypothesis allows us to presume a worst-case of influencing factors, leading the perception performance to exceed the system's boundary of robustness and fault-tolerant ability. While the reasonability of the worst-case can be assessed by the possibility of occurrence. 

\section{Ontology-based triggering condition generation method}
In the previous sections, we have introduced an analysis framework of triggering conditions. From the perspective of the SOTIF, to support the design and safety argumentation of AVs, it is essential to systematically and comprehensively analyze triggering conditions to prevent the omission of critical triggering conditions in the development stage. Therefore, we propose a method to generate potential triggering conditions based on ontology. We defined two ontologies: triggering sources ontology and perception stages ontology. For specific concepts composing the ontologies, their properties and relationships between each other were defined. Based on these two ontologies, triggering conditions can be generated according to the propagation chain of events model. 

\subsection{Ontology construction}
\subsubsection{Ontology of triggering sources}
\begin{figure}[!htpb]
\centering
\includegraphics[width=3.3in]{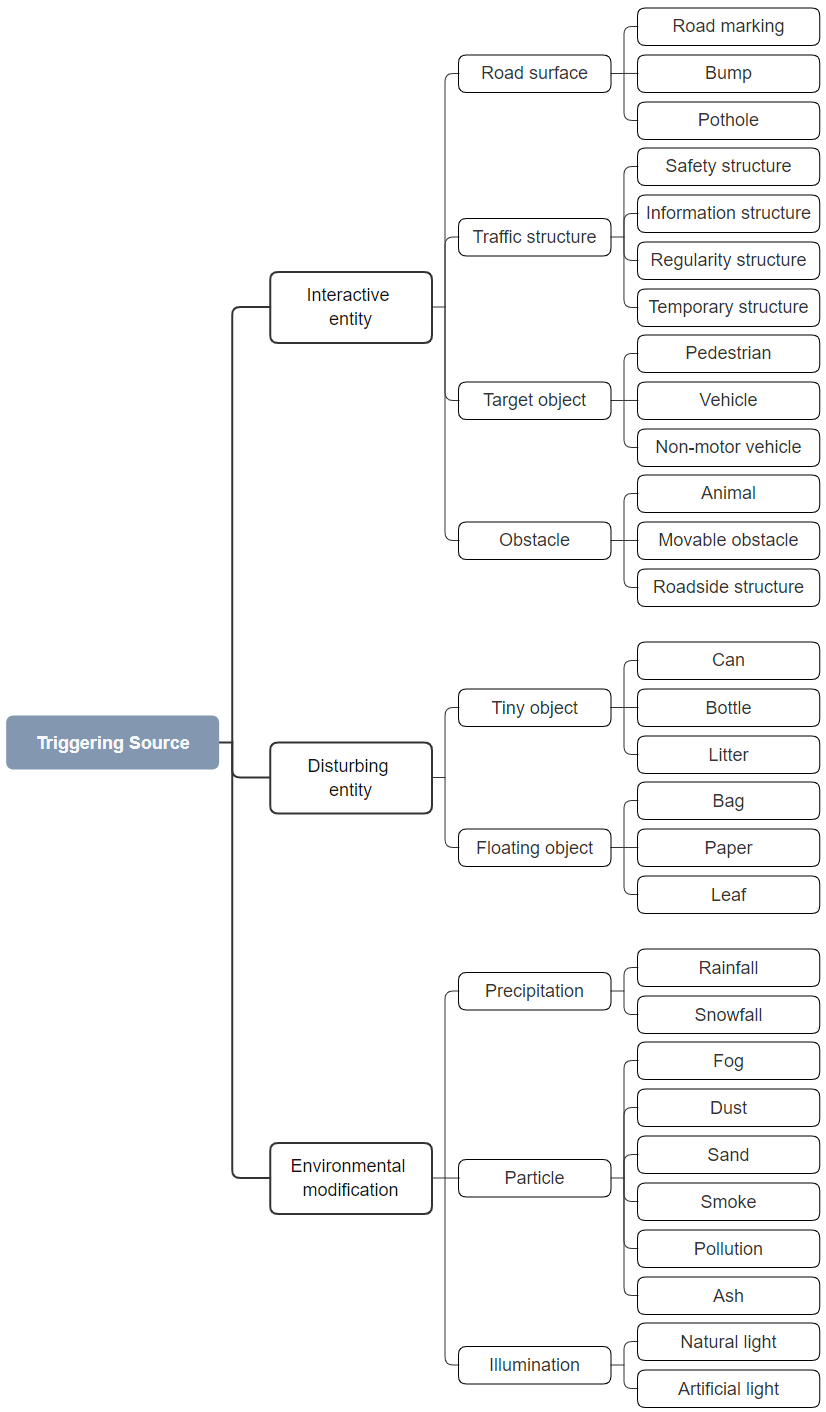}
\caption{Ontology of triggering sources}
\label{Figure 5}
\end{figure}

Triggering sources ontology consists of elements in the traffic environment to determine which elements with specific attributes in the scenario may form triggering conditions. The ontology is organized according to the interactive relationship between AVs and environment elements. This ontology involves three concepts: Interactive entity, Disturbing entity, and Environmental modification. The concept Interactive entity refers to the elements with specific physical attributes that interact with AV in dynamic driving. The concept Disturbing entity refers to elements whose volume and mass can be ignored and will not collide with AVs. The classification of these two concepts is mainly considering the different ways for AVs to deal with them. AVs must correctly perceive and classify interactive entities to assure the safety of driving. On the contrary, AVs should ignore disturbing entities in perception and not be disturbed by them. The concept Environmental modification refers to the atmospheric conditions related to the time of day, weather, and particles, which can change the environmental properties globally. Fig.\ref{Figure 5} shows the concepts in triggering source ontology.

For specific concepts in this ontology, we defined them in terms of properties and instances. The definition of properties mainly involves the physical and cognitive properties related to sensing and recognition. For entity (interactive and disturbing) concepts, their properties include reflectivity-related property, reflection-area-related property, data-generation-related property, and feature-variability-related property. For environmental modification, their properties involve reflectivity-related property and transmittance-related property. Table \ref{table1} and Table \ref{table2} give examples of the concepts \textit{vehicle} and \textit{rainfall}.

\begin{table}[!htbp]
\caption{Definition of concept \textit{Vehicle}}
\label{table1}
\begin{tabular}{|c|l|}
\hline
Concept    & \multicolumn{1}{c|}{\textit{Vehicle}}                                                                                                                                                                                                                                        \\ \hline
Properties & \begin{tabular}[c]{@{}l@{}}\textbf{Reflectivity-related properties}: Surface material, Color,\\ Structure\\ \textbf{Reflection-area-related properties}: Perspective shape \\ \textbf{Data-generation-related property}: Velocity\\ \textbf{feature-variability-related property}: Accessory\end{tabular} \\ \hline
Instances  & Passenger car, Minibus, Bus, Truck, motorcycle                                                                                                                                                                                                                      \\ \hline
\end{tabular}
\end{table}

\begin{table}[!htbp]
\caption{Definition of concept \textit{Rainall}}
\label{table2}
\begin{tabular}{|c|l|}
\hline
Concept                         & \multicolumn{1}{c|}{\textit{Rainall}}                                                                                                                            \\ \hline
Properties                      & \begin{tabular}[c]{@{}l@{}}\textbf{Reflectivity-related properties}: composition, density\\ \textbf{Transmittance-related properties}: density, particle size\end{tabular} \\ \hline
\multicolumn{1}{|l|}{Instances} & Drizzle, Moderate rain, Rain with wind, Sleet                                                                                                            \\ \hline
\end{tabular}
\end{table}

\subsubsection{Ontology of perception stage}
\begin{figure}[!bp]
\centering
\includegraphics[width=3.2in]{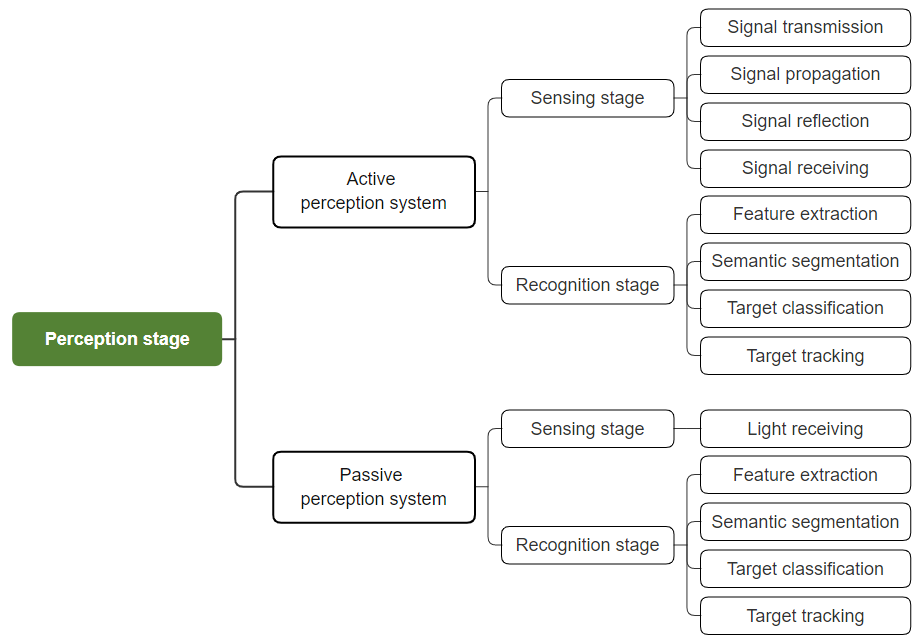}
\caption{Ontology of perception stages}
\label{Figure 6}
\end{figure}

Perception stage Ontology was built to describe the stages in the perception process according to perception systems' working principles, as shown in Fig.\ref{Figure 6}. In this ontology, we firstly defined concepts \textit{Active perception system} and \textit{Passive perception system}. For different perception systems, their perception process is divided into \textit{Sensing stage} and \textit{Recognition stage}. The sensing stage of active perception systems includes four concepts: \textit{Signal transmission}, \textit{Signal propagation}, \textit{Signal reflection}, \textit{Signal receiving}. While for passive perception systems, the sensing stage only involves \textit{light receiving}. The recognition stage of active and passive perception systems consists of \textit{Feature extraction}, \textit{Semantic segmentation}, \textit{Target classification}, and \textit{Target tracking}.

For the Sensing stage concept, their properties were defined considering essential indicators of signal quality. The concepts' properties of the \textit{Active perception system} include \textit{Signal intensity}, \textit{Signal amount}, \textit{Signal noise}. For the \textit{Passive perception system}, the properties include \textit{Brightness}, \textit{Contrast}, and \textit{Purity}. The properties of the \textit{Recognition stage} were defined considering the difficulty to recognize, such as \textit{Variety, Similarity, Contradiction, and Visibility }of features.

\subsection{The relationships of concepts}
\begin{figure}[!htpb]
\centering
\includegraphics[width=3in]{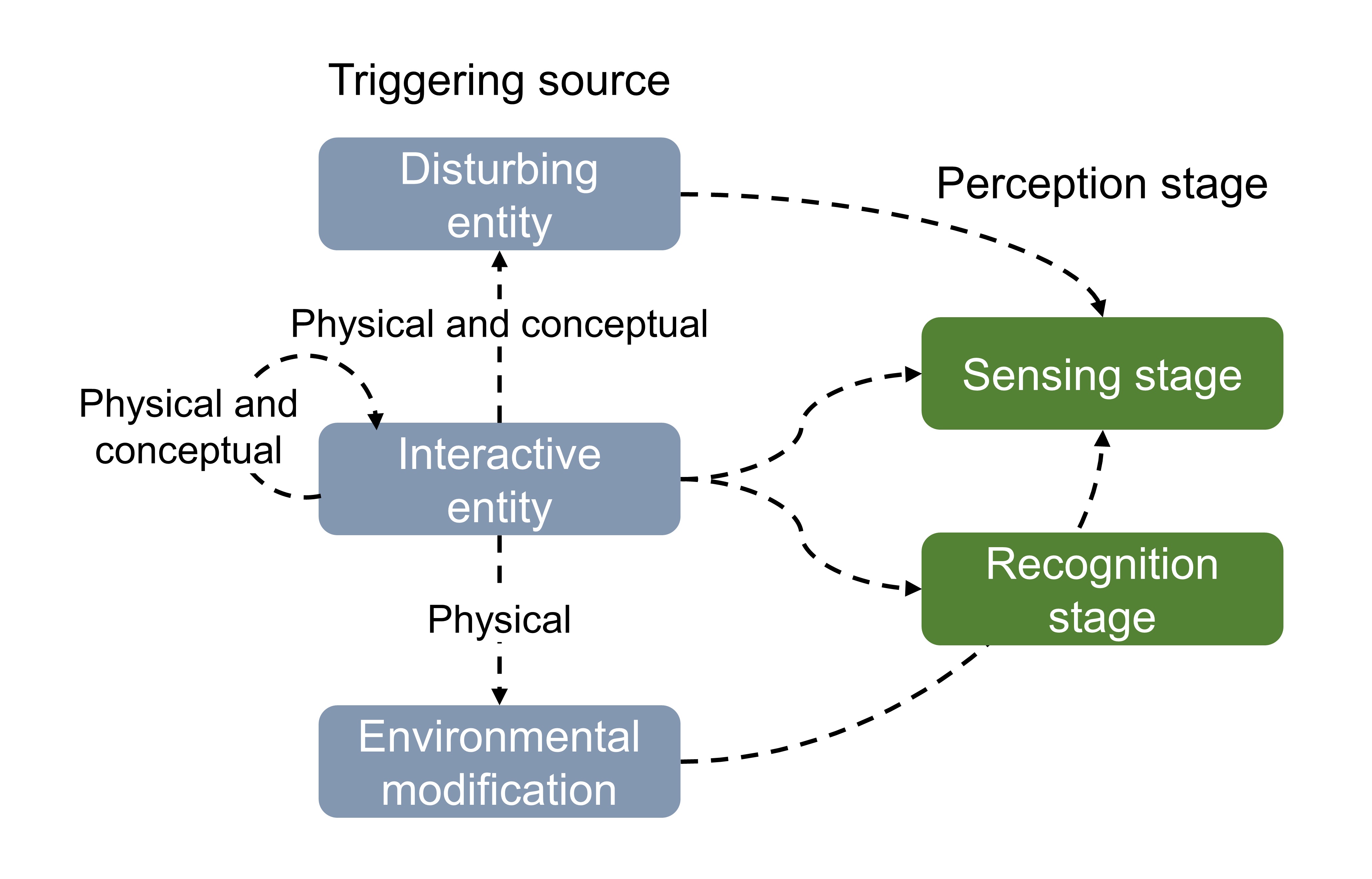}
\caption{relationships structure of triggering source and perception stage}
\label{Figure 7}
\end{figure}

The relationships were derived from physical principles and interactions between different concepts. These relationships were built to reveal the formation mechanism of complicated triggering conditions involving multiple triggering sources. Among triggering sources, relationships were specified with the interactive entity as the center, focusing on describing the influences of the disturbing entity and environmental modification, and the relationships include physical and conceptual ones. In addition, different triggering sources will influence different perception stages. According to section II, we supposed each triggering source would affect the sensing stage, but only interactive entities affect the recognition stage directly. Figure \ref{Figure 7} presents the structure of relationships.

We considered four kinds of relationships among the triggering source concepts from the physical and conceptual perspectives, leading to the changes in interactive entities' properties and features. The specific relationships are as follows.

\begin{itemize}
\item{Physical relationship}
\begin{itemize}
\item[i.]{Spatial position: Relationships to describe the situation where the Interactive entity will be overlayed or occluded by the Disturbing entity. These relationships will lead to changes of the interactive entity's features.}
\item[ii.]{Surface treatment: Relationships to describe the situation where the Interactive entity will be covered or lightened by the Environmental modification. It can change the attribute of the entity's surface, cause difficulties in object detection for sensors.}
\end{itemize}
\item{Conceptual relationship}
\begin{itemize}
\item[i.]{Possess: Some Interactive entities may carry artificial light, obstacles, or tiny objects, influencing the sensing and recognition performance of perception systems. }
\item[ii.]{Cognitive feature: This relationship mainly refers to the feature similarity between Target object, Obstacle, Tiny object, and Floating object. It will challenge the reliability of recognition algorithms. }
\end{itemize}
\end{itemize}

\begin{table*}[!bp]
\centering
\caption{Specified relationships between triggering sources and perception stages}
\label{table3}
\includegraphics[width=6.1in]{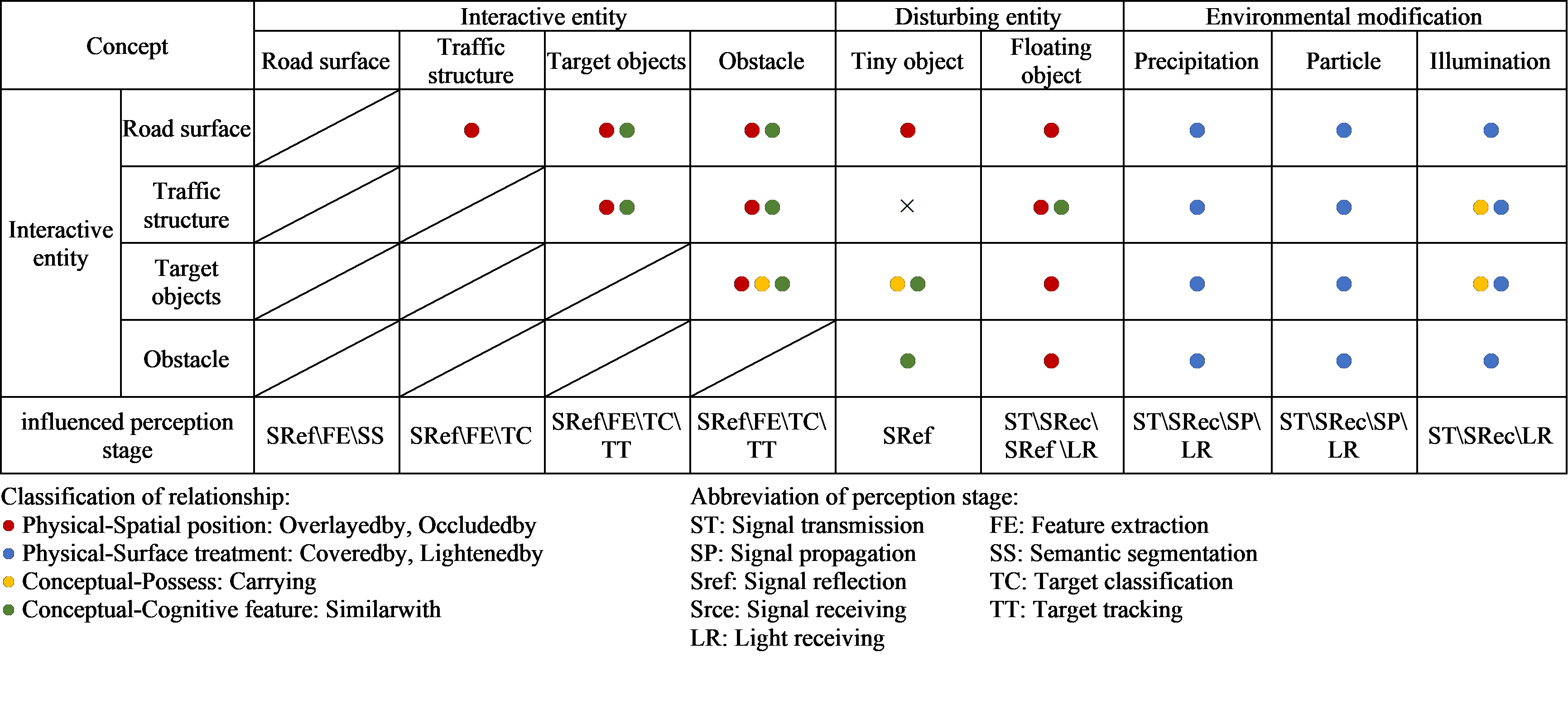}
\end{table*}

For all of these relationships, it is not constrained to consist one of them in a triggering condition. Several relationships may contribute together to trigger perception insufficiencies. For example, the road surface may reflect sunlight strongly due to mirrored surface itself or covered by rainfall water. In this example, surface treatment to the road surface both from precipitation and illumination need be concerned. The complete relationships between different triggering sources are shown in Table \ref{table3}.

The relationships between triggering sources and the perception stages were determined according to entities' properties and the working principle of perception systems. For entities with a certain reflective area, their properties would influence the signal reflection stage. Besides, floating objects may obstruct the sensor and influence the signal transmission and receiving stages. For environmental modification, they can influence the signal propagation stage by changing the medium's attributes. Also, if components of environmental modification cover and obstruct the sensor, it will affect the signal transmission and receiving stages. For instance, the \textit{Signal transmission} and \textit{Signal receiving} of LiDAR can be obstructed if the sensor is covered by water droplets, dust, or plastic bags. The \textit{Signal propagation} also can have interference from rainfall and dust in the air.

\subsection{Triggering condition generation}
Different triggering sources will have different triggering effects on the perception system according to triggering principles. Therefore, every specific triggering source with different properties and its affected perception stages should be analyzed considering all involved relationships. 

We constructed the generation matrix for triggering conditions to cross-analyze the triggering effect by which specific triggering source's properties and the affected perception stages. The generation of triggering conditions consists of three main steps. Table \ref{table4} shows the generating process of triggering conditions for a LiDAR system through the generation matrix.

\textit{Step 1. }Determine the triggering source and the related perception stages from ontology and cross-analyze the triggering effect based on triggering sources and perception stages' properties. Considering the worst-case hypothesis of the triggering source's properties, determine the positive (+) or negative (−) effect on the perception stage and degree of effect.

\textit{Step 2.}Identify possible triggering conditions based on the triggering effect. Besides various external factors, the target distance plays a crucial role in triggering conditions of perception insufficiency. Most of the triggering conditions will reduce the effective detection distance of the sensor significantly. A distant target combined with other triggering conditions will make it harder for the perception system to detect and recognize.

\textit{Step 3.}Assess triggering conditions considering the occurrence probability of worst-case and the criticality of the perception insufficiencies. For the probability, we defined four classes from E1~E4. E1 has the lowest possibility, and E4 has the highest. For the criticality, we also defined four classes from C1~C4 as the same. This assessment is qualitative and gives a relative priority to triggering conditions. 

\begin{table*}[!htpb]
\centering
\caption{Triggering condition analysis of LiDAR formed by the movable obstacle}
\label{table4}
\includegraphics[width=6.1in]{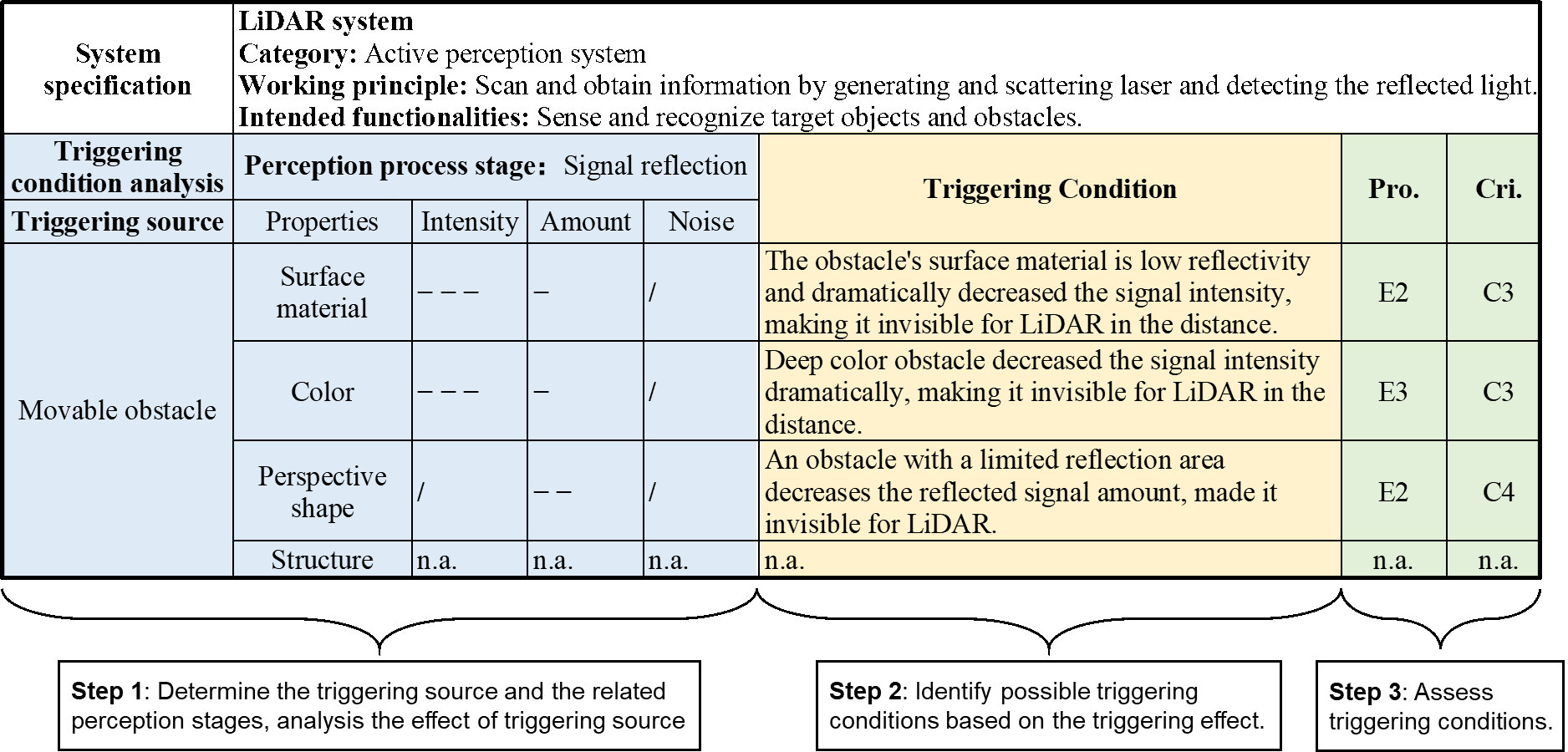}
\begin{threeparttable}
\begin{tablenotes}
\footnotesize
\item {*The properties of triggering source \textit{Movable obstacle} and perception stage \textit{Signal reflection} were cross-analyzed in the table. By assuming a worst-case of \textit{Surface material}, we identified that an extremely low reflective material could decrease the intensity of the reflected signal. Meanwhile, a low reflective material may also reduce the amount of the reflected signal. So, the highest level of negative effect was given for \textit{Surface material} on \textit{Intensity} of \textit{Signal reflection}as "− − −", and the lowest level of negative effect was given on \textit{Amount} of \textit{Signal reflection} as "− ". Other properties were analyzed as the same.}
\end{tablenotes}
\end{threeparttable}
\end{table*}

\begin{table}[bp]
\caption{Functions specification of the autonomous vehicle}
\centering
\begin{tabular}{|l|l|l|}
\hline
\textbf{Functions} & \textbf{Driving Tasks} & \makecell[l]{\textbf{Operational}\\\textbf{Conditions}}\\
\hline
Drive & \makecell[l]{Drive along curbs or with GPS\\signal} & \makecell[l]{Road curbs or\\stable GPS signal}\\
\hline
\makecell[l]{Avoid\\pedestrians }&\makecell[l]{Detect pedestrians ahead, whi-\\stle, and brake to stop, waiting\\for pedestrians to leave} & Pedestrian(s) ahead\\
\hline
\makecell[l]{Avoid\\obstacles} & \makecell[l]{Detect obstacles ahead, and by-\\pass to avoid obstacles}& \makecell[l]{Curbs and obstacles\\ahead}\\
\hline
\makecell[l]{Brake\\to stop }& \makecell[l]{Detect obstacles in close front\\ and brake to avoid obstacles }& \makecell[l]{No curbs but stable\\GPS signals, obsta-\\cles in close front}\\
\hline
\end{tabular}
\label{table5}
\end{table}

\section{Triggering condition tests combined with hazardous events}
In order to validate the method, we analyzed an SAE L3 autonomous vehicle developed by our team and verified its performance as a whole vehicle under specific triggering conditions. According to ISO 21448, it is necessary to verify the AV's safety under hazardous events combined with triggering conditions because the system's performance insufficiencies resulting from triggering conditions will lead to accidents. As for hazardous analysis, existing methods like fault tree analysis (FTA), failure mode and effect analysis (FMEA), and systems-theoretic process analysis (STPA) are applicable. In\cite{ref26}, we have studied the STPA method for hazard analysis and identified 16 hazardous events of this L3 autonomous vehicle. This section analyzes triggering conditions using the proposed method and selects two typical hazardous events to generate test cases. 

The L3 autonomous vehicle is a road sweeper of which the perception system constitutes one front camera and two mechanical rotating LiDAR. The algorithms to identify pedestrians, objects and make decisions are implemented on an Nvidia PX2. The vehicle can plan a path and drive autonomously in a limited-access area like parks and campuses in the autonomous driving mode. During its operation, it can detect and avoid obstacles and pedestrians ahead. The functions under the autonomous mode and their corresponding operational conditions are shown in Table \ref{table5}

Because the vehicle is a road sweeper, the function of driving along curbs and avoiding obstacles near curbs is vital. Furthermore, the vehicle should follow the GPS signal only if there are no curbs, and the corresponding behavior of obstacle avoidance is designed to be conservative as the braking to stop instead of evading. Although the vehicle's driving speed is low, it is still meaningful to take the analysis and test involving triggering conditions to provide evidence for the subsequent improvement of the system.

\subsection{Analysis of triggering conditions}
\subsubsection{Specification of the perception system}
The AV's front camera is used to recognize pedestrians and cyclists, and the LiDAR is used to detect other obstacles and curbs. There is no data fusion for the camera and LiDAR in this system. According to the design requirement of the autonomous road sweeper, the vehicle should operate safely under normal and light rain conditions in the daytime or nighttime with sufficient illumination. Therefore, the perception system should detect and recognize targets reliably under these operational conditions. The detailed information of the perception system is shown in Table\ref{table6}.

\begin{table*}[bp]
\centering
\renewcommand\arraystretch{1.2}
\caption{Specification of the perception system of autonomous vehicle}
\label{table6}
\begin{tabular}{|c|l|l|l|l|} 
\hline
\textbf{\makecell[c]{Sensor\\type}} & \multicolumn{1}{c|}{\textbf{Working principle}}                                                                                                                                                                                                                                                                    & \multicolumn{1}{c|}{\textbf{Perception stages}}                                                                                                                                                                                                                                                                                                   & \multicolumn{1}{c|}{\textbf{Intended functionality}}                                                                                                                                                                                                                                                                & \multicolumn{1}{c|}{\textbf{\makecell[c]{Operational design\\domain}}}                                                                                                                                                                     \\ 
\hline
Camera                                                                                                        & 
\begin{tabular}[c]{@{}l@{}}Receiving light through the lens and forming images\\of the environment by CCD and D/A converter. Then,\\the recognition algorithm working on the images to \\extract features of targets and identify them.\end{tabular}                                                             & 
\begin{tabular}[c]{@{}l@{}}\textbf{Sensing stage}\\\begin{tabular}{@{\labelitemi\hspace{\dimexpr\labelsep+0.5\tabcolsep}}l@{}}Light receiving\end{tabular}\\\textbf{Recognition stage}\\\begin{tabular}{@{\labelitemi\hspace{\dimexpr\labelsep+0.5\tabcolsep}}l@{}}Feature extraction\\Target identification\end{tabular}\end{tabular} & \begin{tabular}[c]{@{}l@{}}\begin{tabular}{@{\labelitemi\hspace{\dimexpr\labelsep+0.5\tabcolsep}}l@{}}Pedestrian recognition~\end{tabular}\\~ ~ and positioning;\\\begin{tabular}{@{\labelitemi\hspace{\dimexpr\labelsep+0.5\tabcolsep}}l@{}}Cyclist recognition~\end{tabular}\\~ ~ and positioning;\end{tabular}          &
\multirow{2}{*}{\begin{tabular}[c]{@{}l@{}}\begin{tabular}{@{\labelitemi\hspace{\dimexpr\labelsep+0.5\tabcolsep}}l@{}}\makecell[l]{Limited access area;}\\\makecell[l]{Daytime without\\ rain;}\\\makecell[l]{Daytime with light\\rain;}\\\makecell[l]{Nighttime with suf-\\ficient illumination;}\end{tabular}\\\end{tabular}}  \\ 

\cline{1-4}
LiDAR                & \begin{tabular}[c]{@{}l@{}}Scanning and obtaining information by scattering laser\\ and detecting the reflected laser. The reflected laser is\\processed to form the points cloud. Then, the recogn-\\ition algorithm working on the points cloud to extract\\features of targets and identify them.\\\end{tabular} & \begin{tabular}[c]{@{}l@{}}\textbf{Sensing stage}\\\begin{tabular}{@{\labelitemi\hspace{\dimexpr\labelsep+0.5\tabcolsep}}l@{}}Signal transmission\\Signal propagation \\Signal reflection \\Signal receiving \end{tabular}\end{tabular}                                                                                                           & \begin{tabular}[c]{@{}l@{}}\begin{tabular}{@{\labelitemi\hspace{\dimexpr\labelsep+0.5\tabcolsep}}l@{}}Curbs detection and\end{tabular}\\~ ~  positioning;\\\begin{tabular}{@{\labelitemi\hspace{\dimexpr\labelsep+0.5\tabcolsep}}l@{}}Obstacles detection\end{tabular}\\~ ~ and positioning;\end{tabular}          &                                                                                                                                                                                                                             \\
\hline
\end{tabular}
\end{table*}
\subsubsection{Triggering condition analysis}
According to the specification of perception systems, we selected the pedestrian, movable obstacle, and roadside structure from the interactive entity, leaf and litter from the disturbing entity, illumination and rain from the environmental modification as the triggering sources. Resulting from these seven triggering sources, we identified 87 triggering conditions in total by applying the steps shown in Table \ref{table4}. Among them, we chose 20 typical triggering conditions, as Table \ref{table7} listed, to test the AV.

\begin{table*}[!htpb]
\caption{List of triggering conditions}
\label{table7}
\centering
\renewcommand\arraystretch{1.2}
\begin{tabular}{|c|c|c|c|c|c|} 
\hline
\textbf{No.} & \textbf{Sensor} & \textbf{Triggering sources}                                                             & \textbf{Properties}                                              & \textbf{Process stage}      & \textbf{Triggering condition}                                                                                                                   \\ 
\hline
T1           & Cam.\textbf{}   & Natural light\textbf{}                                                                  & Light angle\textbf{}                                             & S.-Light receiving\textbf{} & \begin{tabular}[c]{@{}c@{}}A strong sunlight behind the pedestrian at nightfall\textbf{}\end{tabular}                                    \\ 
\hline
T2           & Cam.\textbf{}   & Artificial light\textbf{}                                                               & Light intensity\textbf{}                                         & S.-Light receiving\textbf{} & \begin{tabular}[c]{@{}c@{}}A streetlamp behind the pedestrian at night\textbf{}\end{tabular}                                                   \\ 
\hline
T3           & Cam.            & Leaf                                                                                    & Material                                                         & S.-Light
  receiving        & Fallen
  leaves cover the camera                                                                                                                \\ 
\hline
T4           & Cam.            & Rain                                                                                    & Composition                                                      & S.-Light
  receiving        & Water
  droplets cover the camera                                                                                                               \\ 
\hline
T5           & Cam.            & Pedestrian                                                                              & Accessory                                                        & R.-Classification           & \begin{tabular}[c]{@{}c@{}}A pedestrian carries an umbrella occluding his head\end{tabular}                                                \\ 
\hline
T6           & Cam.            & Pedestrian                                                                              & Accessory                                                        & R.-Classification           & A
  pedestrian who is skateboarding                                                                                                             \\ 
\hline
T7           & Cam.            & Pedestrian                                                                              & Perspective
  shape                                              & R.-Classification           & \begin{tabular}[c]{@{}c@{}}A pedestrian who is squatting on the side of the road\end{tabular}                                                 \\ 
\hline
T8           & Cam.            & Pedestrian                                                                              & Perspective
  shape                                              & R.-Classification           & A
  pedestrian who is sitting by the road                                                                                                       \\ 
\hline
T9           & Cam.            & \begin{tabular}[c]{@{}c@{}}Occludedby(Pedestrian,\\Temporary structure)\end{tabular}    & Perspective
  shape                                              & R.-Classification           & \begin{tabular}[c]{@{}c@{}}A pedestrian whose legs are occluded by traffic barrels\end{tabular}                                               \\ 
\hline
T10          & Cam.            & \begin{tabular}[c]{@{}c@{}}Occludedby (Pedestrian,\\Temporary structure)\end{tabular}   & Perspective
  shape                                              & R.-Classification           & \begin{tabular}[c]{@{}c@{}}A pedestrian whose legs are occluded by traffic \\barrels wears similar color clothes with the barrels\end{tabular}  \\ 
\hline
T11          & Cam.            & \begin{tabular}[c]{@{}c@{}}Occludedby (Pedestrian, \\Regularity structure)\end{tabular} & Perspective
  shape                                              & R.-Classification           & \begin{tabular}[c]{@{}c@{}}A pedestrian whose torso is occluded by traffic signs\end{tabular}                                                 \\ 
\hline
T12          & Cam.            & \begin{tabular}[c]{@{}c@{}}Occludedby (Pedestrian, \\Regularity structure)\end{tabular} & Perspective
  shape                                              & R.-Classification           & \begin{tabular}[c]{@{}c@{}}A pedestrian whose torso and legs are occluded by \\traffic signs\end{tabular}                                       \\ 
\hline
T13          & Cam.            & \begin{tabular}[c]{@{}c@{}}Similarwith(Pedestrian,\\~Movable obstacle)\end{tabular}     & \begin{tabular}[c]{@{}c@{}}Perspective\\shape/Color\end{tabular} & R.-Classification           & A
  billboard with a human picture on it                                                                                                        \\ 
\hline
T14          & LiD.            & Leaf                                                                                    & Perspective
  shape                                              & S.- Laser
  transmission    & Fallen
  leaves cover the LiDAR                                                                                                                 \\ 
\hline
T15          & LiD.            & Rain                                                                                    & Perspective
  shape                                              & S.- Laser
  transmission    & Water
  droplets cover the LiDAR                                                                                                                \\ 
\hline
T16          & LiD.            & Litter                                                                                  & Perspective
  shape                                              & S.-Signal
  reflection      & A
  pile of rubbish on the side of the road                                                                                                     \\ 
\hline
T17          & LiD.            & Movable
  obstacle                                                                      & Perspective
  shape                                              & S.-Signal
  reflection      & A
  thin carton on the road                                                                                                                     \\ 
\hline
T18          & LiD.            & Movable
  obstacle                                                                      & Material                                                         & S.-Signal
  reflection      & \begin{tabular}[c]{@{}c@{}}A carton made by low reflective material on the road\end{tabular}                                                  \\ 
\hline
T19          & LiD.            & Roadside
  structure                                                                    & Perspective
  shape                                              & S.-Signal
  reflection      & A
  thin rod sticks out from the roadside structure                                                                                             \\ 
\hline
T20          & LiD.            & \begin{tabular}[c]{@{}c@{}}Occludedby (Road surface,\\Floating object)\end{tabular}     & Material                                                         & S.-Signal
  reflection      & A
  cloth with low reflective covers up the curbs                                                                                               \\ 
\hline
\multicolumn{6}{r}{*\textit{S.} represents the \textit{Sensing stage}, \textit{R.}represents the \textit{Recognition stage}.}                                                                                                                                                                                                                                              
\end{tabular}
\end{table*}

\subsection{Test cases generation}
\begin{figure}[tp]
\centering
\includegraphics[width=3in]{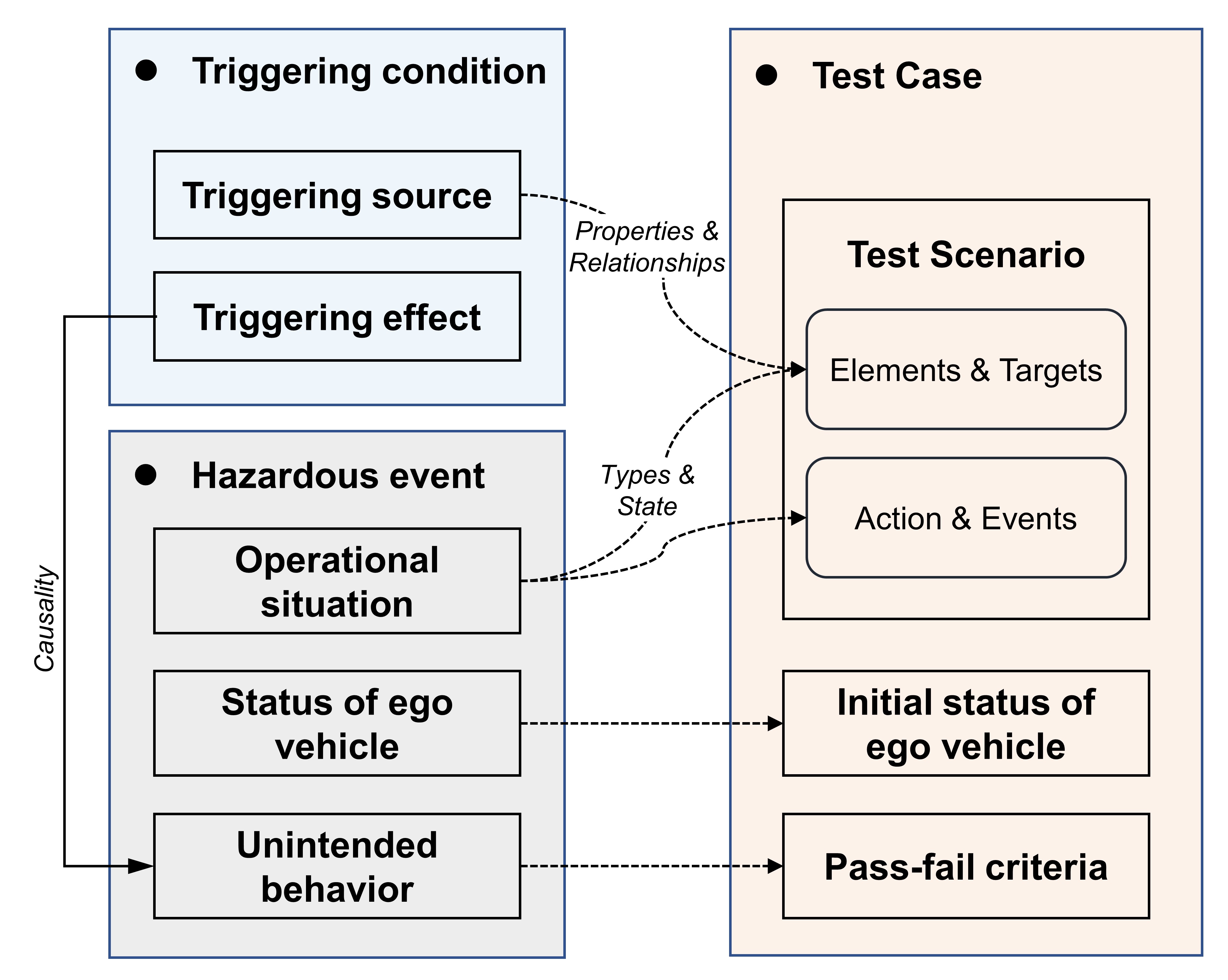}
\caption{Generation framework for test cases combining with triggering conditions}
\label{Figure 8}
\end{figure}

In order to verify the vehicle's behavior under potential triggering conditions, we generated test cases and implemented a field test. In\cite{ref12}, we proposed a testing case generation framework based on triggering conditions and hazardous events, as Fig.\ref{Figure 8} represented. First of all, the triggering condition combines with the operational situation to constitute the test scenario. Meanwhile, the triggering condition will directly cause unintended behavior of the AV, and the pass-fail criteria are defined according to the AV's behaviors. 

\begin{figure}[tp]
\centering
\includegraphics[width=3in]{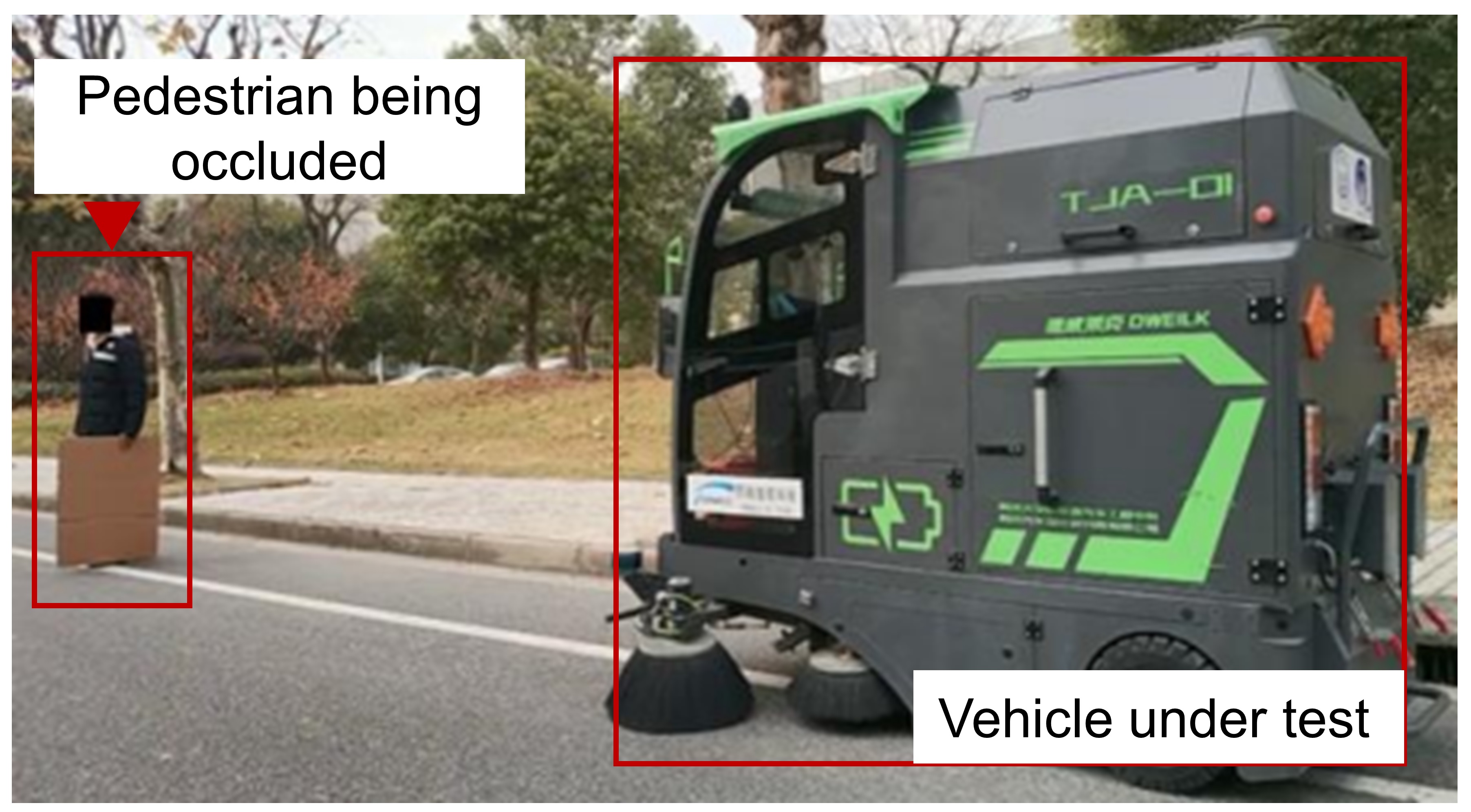}
\caption{Field test of the L3 autonomous vehicle}
\label{Figure 9}
\end{figure}

We selected two specific hazardous events as examples according to the AV's function specification. For each hazardous event, the operational situation, status of the ego vehicle, and unintended behavior are shown in Table \ref{table8}. The triggering conditions identified in Table \ref{table7} are implanted in hazardous events as supplementary information of test scenarios. If the perception system cannot recognize the front object timely and accurately, the vehicle may be at risk of collision.

\begin{table}[!htpb]
\caption{Specification of hazardous events in obstacle and pedestrian avoidance functions}
\label{table8}
\centering
\begin{tabular}{|c|lll|}
\hline
\multirow{3}{*}{\textbf{Function}} & \multicolumn{3}{c|}{\textbf{Hazardous event}} \\ 
\cline{2-4} 
 &
\multicolumn{1}{c|}{\textit{\makecell[c]{Operational\\situation}}}                                                                         & \multicolumn{1}{c|}{\textit{\makecell[c]{Status of\\ego vehicle}}}                                                         & \multicolumn{1}{c|}{\textit{\makecell[c]{Unintended\\behavior}}}                                              \\ \hline
\makecell[c]{Obstacle\\avoidance}        & \multicolumn{1}{l|}{\begin{tabular}[c]{@{}l@{}}With curbs and\\ stable GPS sig-\\nal, obstacles\\ahead\end{tabular}}   & \multicolumn{1}{l|}{\begin{tabular}[c]{@{}l@{}}Driving straight\\along the road-\\side\end{tabular}} & \begin{tabular}[c]{@{}l@{}}Continue driving\\ with no bypass\\maneuver\end{tabular} \\ \hline
\makecell[c]{Pedestrian\\avoidance}      & \multicolumn{1}{l|}{\begin{tabular}[c]{@{}l@{}}With curbs and\\ stable GPS sig-\\nal, pedestrians\\ahead\end{tabular}} & \multicolumn{1}{l|}{\begin{tabular}[c]{@{}l@{}}Driving straight\\along the road-\\side\end{tabular}} & \begin{tabular}[c]{@{}l@{}}Continue driving\\ with no brake\end{tabular}              \\ \hline
\end{tabular}
\end{table}

\begin{figure*}[!htpb]
\centering
\includegraphics[width=6.5in]{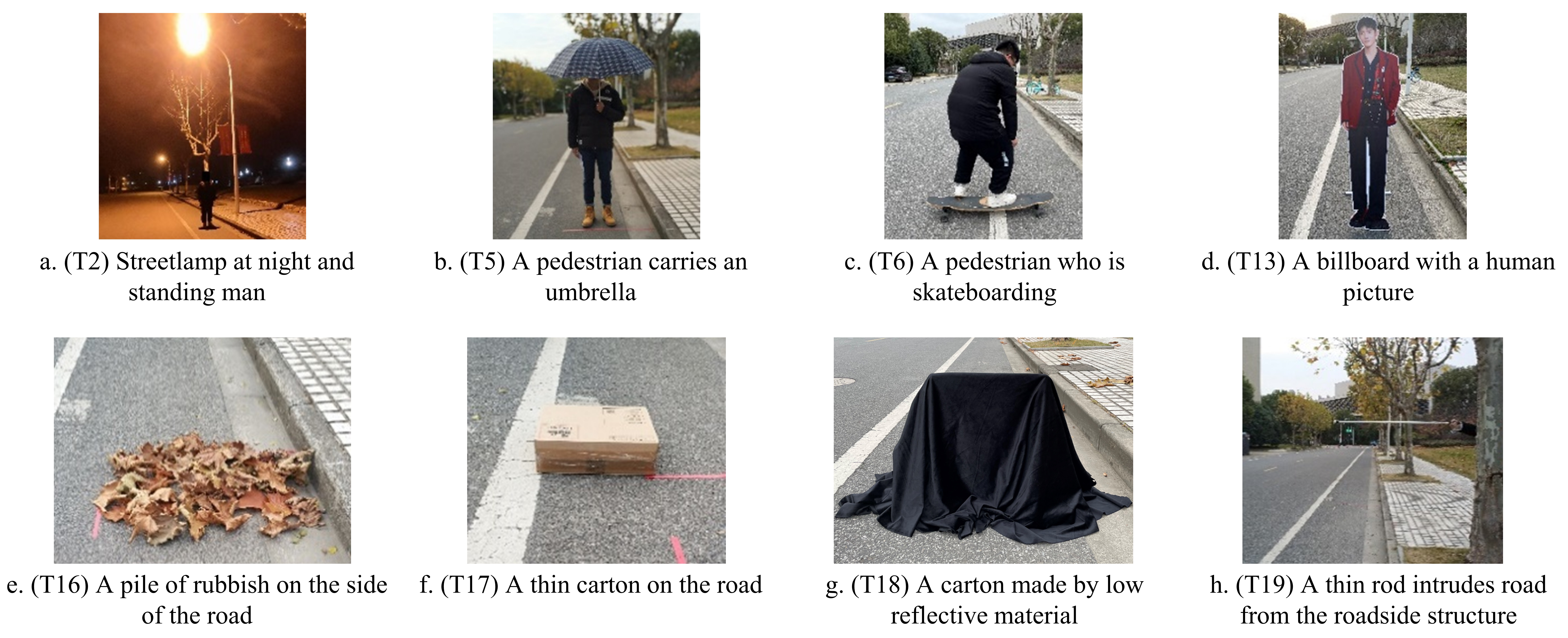}
\caption{Instances of triggering sources used in tests}
\label{Figure 10}
\end{figure*}

The tests were executed on an inner road of a campus (Fig.\ref{Figure 9}). Before the triggering conditions tests, a controlled test with typical obstacles and pedestrians was executed three times to verify the functions' validity. In tests, the AV's driving data, including velocity, heading, location, were recorded synchronously with the test process video. The triggering sources were constructed using real things as requirements. Some targets and triggering sources are shown in Fig.\ref{Figure 10}.

\subsection{Test results analysis}
According to the test results of the above 20 test cases, 15 triggering conditions were verified to be effective and led to the vehicle's unexpected behaviors. Among these 15 triggering conditions, two of them caused unreliable recognition results and led to a collision with the target. Six triggering conditions caused wrong classification results and induced risky behavior of the vehicle in pedestrian avoidance occasions. Besides the above nine triggering conditions that formed actual hazardous events accordingly, five triggering conditions caused unstable perception results, making the vehicle hesitate to brake when it encountered a pedestrian. The verified hazardous events and the corresponding triggering conditions are shown in Table \ref{table9}.

\begin{table}[bp]
\centering
\renewcommand\arraystretch{1.2}
\caption{Effective triggering conditions and resulted hazardous events in the tests}
\label{table9}
\begin{tabular}{|c|l|l|} 
\hline
\textbf{No.} & \multicolumn{1}{c|}{\textbf{Hazardous events in tests}}                                            & \multicolumn{1}{c|}{\textbf{Triggering conditions}}  \\ 
\hline
1            & \begin{tabular}[c]{@{}l@{}}Near collision with targets (intervened\\by safety driver)\end{tabular} & T15/T20                                              \\ 
\hline
2            & \begin{tabular}[c]{@{}l@{}}Risky behavior due to the wrong\\classification\end{tabular}            & T1/T3/T6/T7/T9/T12                                   \\ 
\hline
3            & \begin{tabular}[c]{@{}l@{}}Hesitate to brake when encountering\\pedestrians\end{tabular}           & T4/T5/T8/T10/T11                                     \\ 
\hline
4            & Unintended behaviors without
  hazard                                                              & T13/T16                                              \\
\hline
\end{tabular}
\end{table}

In triggering condition T15, the LiDAR's front shell was covered by water droplets, refracting laser beams and preventing the sensor from detecting the target accurately. In triggering condition T20, the LiDAR did not detect a thin rod intruding the road from a roadside structure. Since the LiDAR has only 16 channels and was installed at a low position to detect curbs, its vertical angular resolution and field of view (FOV) are insufficient to detect thin obstacles at a high position. Besides, the above results demonstrate that obstacle detection relies on LiDAR individually. When the LiDAR's perception is unreliable, such collisions are very likely to occur in an actual operation without redundant sensors. In triggering conditions T1, T6, T7, T9, and T12, which involve pedestrian recognition, the camera cannot identify pedestrians due to the unsatisfying conditions of pedestrian features. While the vehicle encountered a pedestrian, the LiDAR detected the pedestrian as a static obstacle, thus the vehicle executed an avoidance maneuver rather than stop and wait. In this circumstance, the vehicle may collide with the moving pedestrian due to a limited ability to deal with moving targets. In triggering conditions T3, the camera was partially covered by fallen leaves, producing incomplete front view images. Therefore, the camera can no longer recognize pedestrians, but the vehicle still drives normally. Meanwhile, the LiDAR detected the pedestrian as a static obstacle, again the vehicle tried to avoid it. Because the tested vehicle does not provide an effective method to monitor the sensor's availability, it still performs the driving function when the sensor goes wrong seriously, which is significantly risky. For triggering conditions that led to hesitation in the pedestrian avoidance, they resulted in an unstable identification of the distant pedestrian. The vehicle would take an avoidance first and brake to stop when nearing the pedestrian or behave as a stop-go-stop. 

The test results prove that the triggering conditions identified through the method affect the perception system significantly. The method can systematically generate triggering conditions to support the testing and verification process of the system's safety.

\section{Discussion}
The method proposed in this paper provides a systematic way to identify triggering conditions in a semantic manner. By applying the method, possible triggering conditions can be specified from their properties and characteristics. The method provides inferences for which triggering sources can potentially affect the perception system but cannot guarantee the validity of every triggering condition. In order to support the verification and validation progress combining scenario tests, there are still several problems that need to be concerned. 

\begin{itemize}
\item{Quantification of triggering conditions}
\end{itemize}

The quantification of semantic triggering conditions for the construction of concrete scenarios is necessary from the perspective of testing. In order to quantify the triggering conditions, more detailed information about the perception system is needed, and proper metrics for triggering conditions must be determined. The determination of metrics should consider the triggering source's properties and the principle of the perception insufficiency to be triggered. Besides, more laboratory and field tests should be done to study the quantitative relationship between the triggering conditions and perception insufficiencies.
\begin{itemize}
\item{Assessment of triggering conditions}
\end{itemize}

Another critical issue is to determine which triggering conditions should be tested in the verification progress. Our method can generate various triggering conditions according to analysis requirements, but not all of them are worth attention due to an extremely low probability of occurrence or rare impact on perception systems. So, it needs a more reliable approach to assess the identified triggering conditions. Data collection through field tests focusing on the perception system's performance is a practical approach to supporting the assessment. Based on the triggering sources ontology or other elements structure, related elements should be labeled in the database to study the possibility of occurrence statistically.

\section{Conclusion}
In this paper, we proposed an ontology-based analysis method of triggering conditions. The triggering source ontology and perception stage ontology were built according to working principle analysis of passive and active perception systems. We specified the relationship between concepts to represent the possible pattern of triggering conditions. Finally, an identification method of triggering conditions based on the ontology and relationships was introduced to generate and assess possible triggering conditions. We verified the proposed method on an L3 autonomous vehicle. Eighty-seven triggering conditions were identified considering the AV's functions and perception system composition. Twenty of these triggering conditions were tested on the field, combined with scenarios generated through hazard analysis, and 15 triggering conditions were verified to be effective. Among the 15 effective triggering conditions, two triggering conditions led the vehicle to collide with the target due to unreliable recognition results. Six triggering conditions induced risky vehicle behavior due to wrong classification results when avoiding a pedestrian. The test results proved the validity of the method.

The method proposed provides a practical way to support identifying triggering conditions.   Compared with the methods that depend on expertise and experience, this method gives a more formal and structured way to describe triggering sources and identifies triggering conditions more wholly and systematically. In the future study, we will implement more tests to study the quantitative relationship between the triggering conditions and perception insufficiencies. Besides, a more reliable assessment of triggering conditions is also vital in future work.


\begin{thebibliography}{1}
\bibliographystyle{IEEEtran}

\bibitem{ref1}
Road Vehicles-Functional Safety, ISO 26262, International Organization for Standardization, Geneva, Switzerland, 2011.

\bibitem{ref2}
Road Vehicles-Safety of the intended functionality, ISO/PAS 21448, International Organization for Standardization, Geneva, Switzerland, 2019.

\bibitem{ref3}
K. Czarnecki, and R. Salay, "Towards a framework to manage perceptual uncertainty for safe automated driving," in \textit{Proc. Int. Conf. Comput. Saf., Rel., Secur.}, Västerås, Sweden, 2018, pp. 439--445.

\bibitem{ref4}
L. Ren, H.-L. Yin, W.-C Ge, and Q. Meng, "Environment influences on uncertainty of object detection for automated driving systems," in \textit{12th Int. Cong. Imag. Signal Process., BioMed. Eng. Informatics}, Suzhou, China, 2019, pp. 1--5.

\bibitem{ref5}
S. Hasirlioglu, I. Doric, A. Kamann, and A. Riener, "Reproducible fog simulation for testing automotive surround sensors," in {\it{IEEE 85th Veh. Tech. Conf.}}, Sydney, NSW, Australia, 2017, pp. 1--7.

\bibitem{ref6}
S. Hasirlioglu, A. Kamann, I. Doric, and T. Brandmeier, "Test methodology for rain influence on automotive surround sensors," in {\it{IEEE 19th Int. Conf. Intell. Transp. Syst.}}, Rio de Janeiro, Brazil, pp. 2242--2247.

\bibitem{ref7}
L. Peng, H. Wang, and J. Li, "Uncertainty evaluation of object detection algorithms for autonomous vehicles," \textit{Automot. Innov.}, vol. 4, no. 3, pp. 241--252, 2021.

\bibitem{ref8}
P. Rosenberger, M. Holder, M. Zirulnik, and H. Winner, "Analysis of real world sensor behavior for rising fidelity of physically based lidar sensor models," in \textit{2018 IEEE Intell. Veh. Symp. }, Changshu, China, 2018, pp. 611--616.

\bibitem{ref9}
A. Filgueira, H. González-Jorge, S. Lagüela, L. Díaz-Vilariño, and P. Arias, "Quantifying the influence of rain in LiDAR performance," \textit{Measurement}, vol. 95, pp. 143--148, 2017.

\bibitem{ref10}
D. McKnight, and R. Miles, "Impact of reduced visibility conditions on laser based DP sensors," in \textit{Dyna. Posi. Conf.}, Houston, Tex, USA, 2014, pp. 1--9.

\bibitem{ref11}
X.-Y. Wu, H.-L. Meng, X.-Y. Xing, and J.-Y. Chen, "Edge test case generation of automatic driving system," \textit{J. Tongji Univ. (Natu. Sci.)}, vol. 46, pp. 111--115, 2018.

\bibitem{ref12}
A. Huang, X.-Y. Xing, T.-R. Zhou, and J.-Y. Chen, "A safety analysis and verification framework for autonomous vehicles based on the identification of triggering events," in \textit{Int. Automot. Secur., Saf. Tes. Cong.}, Shanghai, China, 2020, pp. 1--8.

\bibitem{ref13}
H. W. Heinrich, \textit{Industrial Accident Prevention. A Scientific Approach}, 2nd ed. London, U.K., McGraw-Hill Book Company, Inc., 1941.

\bibitem{ref14}
K. Poland, M. P. McKay, D. Bruce, and E. Becic, "Fatal crash between a car operating with automated control systems and a tractor-semitrailer truck," \textit{Traffic Inj. Prevention }, vol. 19, no. 2, pp. S153--S156, 2018. 

\bibitem{ref15}
Y. Li, and J. Ibanez-Guzman, "Lidar for autonomous driving: The principles, challenges, and trends for automotive lidar and perception systems," \textit{IEEE Signal Process. Mag.}, vol. 37, no. 4, pp. 50--61, 2020.

\bibitem{ref16}
X.-L. Li, Y. Zhou, and B.-C. Hua, "Study of a multi-beam LiDAR perception assessment model for real-time autonomous driving," \textit{IEEE Trans. Instrum. Meas.}, vol. 70, pp. 1--15, 2021.

\bibitem{ref17}
S. Yang, W.-W. Deng, Z.-Y. Liu, and Y. Wang, "Analysis of illumination condition effect on vehicle detection in photo-realistic virtual world," in \textit{Intell. Connected Veh. Sympo.}, Kunshan, China, 2017, pp. 1--6.

\bibitem{ref18}
A. Hanif, A. B. Mansoor, and A. S. Imran, "Performance analysis of vehicle detection techniques: a concise survey," in \textit{World Conf. Inf. Syst. Tech.}, Naples, Italy, 2018, pp. 491--500.

\bibitem{ref19}
T. Fersch, A. Buhmann, A. Koelpin, and R. Weigel, "The influence of rain on small aperture LiDAR sensors," in \textit{2016 Ger. Microw. Conf. }, Bochum, Germany, 2016, pp. 84--87.

\bibitem{ref20}
R. Heinzler, P. Schindler, J. Seekircher, W. Ritter, and W. Stork, "Weather influence and classification with automotive LiDAR sensors," in \textit{2019 IEEE intell. veh. symp.}, Paris, France, 2019, pp. 1527--1534.

\bibitem{ref21}
J. Wojtanowski, M. Zygmunt, M. Kaszczuk, Z. Mierczyk, and M. Muzal, "Comparison of 905 nm and 1550 nm semiconductor laser rangefinders’ performance deterioration due to adverse environmental conditions," \textit{Opto-Electron. Rev.}, vol. 22, no. 3, pp. 183--190, 2014.

\bibitem{ref22}
L.-S. Hou, H.-L. Yin, W.-C. Ge, and Q. Meng, "High-order uncertainty of LiDAR for environment perception in automated driving systems," in \textit{12th Int. Cong. Imag. Signal Process., BioMed. Eng. Informatics}, Suzhou, China, 2019, pp. 1-5.

\bibitem{ref23}
R. H. Asshofer, M. Spies, and H. Spies. "Influences of weather phenomena on automotive laser radar systems," \textit{Adv. Radio Sci. }, vol. 9, no. B.2 pp. 49--60, 2011. 

\bibitem{ref24}
T. Y. Nakajima, T. Imai, O. Uchino, and T. Nagai, "Influence of daylight and noise current on cloud and aerosol observations by spaceborne elastic scattering lidar," \textit{Appl. Opt.}, vol. 38, no. 24, pp. 5218--5228, 1999.

\bibitem{ref25}
W.-H. Li, H.-Y. Ni, Y. Wang, B. Fu, P.-X. Liu, and S.-J. Wang, "Detection of partially occluded pedestrians by an enhanced cascade detector," \textit{IET Intell. Transp. Syst.}, vol. 8, no. 7, pp. 621--630, 2014.

\bibitem{ref26}
J.-Y. Chen, T.-R. Zhou, X.-Y. Xing, and L. Xiong, "Research on safety analysis method for autonomous vehicles based on STPA." \textit{Automob. Tech.}, vol. 12, pp. 1--5, 2019.

\end{thebibliography}
\end{document}